\documentclass{llncs} 

\usepackage{amssymb}
\usepackage{textcomp}
\usepackage{comment}
\usepackage{color}
\usepackage{graphicx}
\usepackage{amsmath}

\newcommand{\Beginproof}{{\em Proof.}  }
\newcommand{\Endproof}{\hfill$\Box$\\}

\newcommand{\ket}[1]{|#1\rangle}

\newtheorem{THM}{Theorem}

\newtheorem{CR}[THM]{Corollary}

\begin{document}

\title{Reordering Method and Hierarchies for Quantum and Classical Ordered Binary Decision Diagrams}
\author{Kamil~Khadiev$^{1,2,}$\thanks{Partially supported by ERC Advanced Grant MQC. The work is performed according to the Russian Government Program of Competitive Growth of Kazan Federal University}\and Aliya~Khadieva$^2$}

\institute{ University of Latvia, Riga, Latvia      
       \and       
Kazan Federal University, Kazan, Russia      
                \\ \email{kamilhadi@gmail.com, aliya.khadi@gmail.com } 
}

\maketitle

\begin{abstract}
We consider Quantum OBDD model. It is restricted version of read-once Quantum Branching Programs, with respect to ``width'' complexity. It is known that maximal complexity gap between deterministic and quantum model is exponential. But there are few examples of such functions. We present method (called ``reordering''), which allows to build Boolean function $g$ from Boolean Function $f$, such that if for $f$ we have gap between quantum and deterministic OBDD complexity for natural order of variables, then we have almost the same gap for function $g$, but for any order. Using it we construct the total function $REQ$ which deterministic OBDD complexity is $2^{\Omega(n/\log n)}$ and present quantum OBDD of width $O(n^2)$. It is bigger gap for explicit function that was known before for OBDD of width more than linear.
Using this result we prove the width hierarchy for complexity classes of Boolean functions for quantum OBDDs. 

Additionally, we prove the width hierarchy for complexity classes of Boolean functions for bounded error probabilistic OBDDs. And using  ``reordering'' method we extend a hierarchy for $k$-OBDD of polynomial size, for $k=o(n/\log^3n)$. Moreover, we proved a similar hierarchy for bounded error probabilistic $k$-OBDD.  And for deterministic and probabilistic $k$-OBDDs of superpolynomial and subexponential size.

\end{abstract}

\section{Introduction}
Branching programs are one of the well known models of computation. These models have been shown useful in a variety of domains, such as hardware verification, model checking, and other CAD applications (see for example the book by I. Wegener \cite{Weg00}). It is known that the class of Boolean  functions computed by polynomial size branching programs coincide with the class of functions computed by non-uniform log-space machines.

One of important restrictive branching programs are oblivious read once branching programs, also known as Ordered Binary Decision Diagrams (OBDD) \cite{Weg00}. It is a good model of data streaming algorithms. These algorithms are actively used in industry, because of rapidly increasing of size of data which should be processed by programs. Since a length of an OBDD is at most linear (in the length of the input), the main complexity measure is ``width'', analog of size for automata.
And it can be seen as nonuniform automata (see for example \cite{ag05}).  
In the last decades quantum OBDDs came into play \cite{agk01}, \cite{nhk00}, \cite{SS05},\cite{s06}.

In 2005 F. Ablayev, A. Gainutdinova, M. Karpinski, C. Moore and C. Pollett \cite{agkmp2005} proved that the gap between width of quantum and deterministic OBDD is at most exponential. They showed that this bound can be reached for $MOD_p$ function that takes the value $1$ on input shuch that number of $1$s by modulo $p$ is $0$. Authors presented quantum OBDD of width $O(\log p)$ for this function (another quantum OBDD of same width is presented in \cite{av2008}) and proved that any deterministic OBDD has width at least $p$.  However explicit function $MOD_p$ presents a gap for OBDD of at most linear width. For bigger width it was shown that Boolean function $PERM$ has not deterministic OBDD of width less than $2^{\sqrt{n}/2}/(\sqrt{n}/2)^{3/2}$ \cite{kmw91} and M. Sauerhoff and D. Sieling \cite{ss2005} constructed quantum OBDD of width $O(n^2\log n)$. F. Ablayev, A. Khasianov and A. Vasiliev \cite{akv2008} presented the quantum OBDD of width $O(n\log n)$ for $PERM$. Let us note that for partial functions the gap between widths of quantum and deterministic OBDDs can be more than exponential \cite{agky14}, \cite{g15}, \cite{agky16}.

Better difference between quantum and deterministic complexity was proven in \cite{akv2008} for Equality function. But there authors had exponential gap only for natural order of variables. They presented the quantum OBDD of width $O(n)$, which is based on quantum fingerprinting technique, at the same time any deterministic OBDD has width at least $2^{n/2}$ for natural order. But if we consider any order, then we can construct deterministic OBDD of constant width. 

Changing the order is one of the main issues for proving lower bound on width for OBDD.  We present a technique that allows to build Boolean Function $g$ from Boolean function $f$. We consider $f$ such that any deterministic OBDD with natural order for the function  has width at least $d(n)$ and we can construct quantum OBDD of width $w(n)$. In that case we present quantum OBDD of width $O(w(n/\log n)\cdot n/\log n)$ for function $g$ and any deterministic OBDD has width at least $d(O(n/\log n))$. It means that if difference between quantum OBDD complexity of function $f$ and deterministic OBDD complexity for natural order is exponential, then we have almost exponential difference for function $g$. We called this method ``reordering''. And idea is based on adding addresses of variables to input. Similar idea was used in \cite{k15}.

Then, we present Boolean function {\em Reordered Equality} ($REQ$), it is modification of  Equality function \cite{akv2008}. We apply the main ideas of {\em reordering} and prove that for  $REQ$ deterministic OBDD  has width  $2^{\Omega(n/log n)}$ and  bounded error quantum OBDD has width $O(n^2/\log^2 n)$.  This gap between deterministic and quantum complexity is better than for $PERM$ function. And it has advantage over results on $EQ$, because we prove distance for any order of input variables. And in comparing to $MOD_p$, we can show a distance for bigger width.

Using complexity properties of $MOD_p$ function, $REQ$ function and {\em Mixed weighted sum function} ($MWS_n$)\cite{s2005} we prove the width hierarchy (not tight) for classes of Boolean functions computed by bounded error quantum OBDD. The hierarchy is separated to three cases: first one is for width less than $\log n$, and for this case we prove hierarchy with small gap between width-parameters of classes. Second one is for width less than $n$ and bigger gap. And third one is for width less than $2^{O(n)}$ and here gap is the biggest.
Similar hierarchy is already known for deterministic and nondeterministic OBDD \cite{agky14}, \cite{agky16}, for deterministic $k$-OBDD \cite{k15}. And we  present not tight width hierarchy for bounded error probabilistic OBDD in the paper.

Forth group of results is extending the hierarchy for deterministic and bounded error probabilistic $k$-OBDD of polynomial size. The known tight hierarchy for classes of Boolean functions that computed by a deterministic $k$-OBDD of polynomial size is result of B. Bollig, M. Sauerhoff, D. Sieling, and I. Wegener \cite{bssw96}. They proved that P-$(k-1)$OBDD $\subsetneq$ P-$k$OBDD for $k=o(\sqrt{n}\log^{3/2}n)$. It is known extension of this hierarchy for $k=o(n/\log^2 n)$ in papers \cite{K16}, \cite{ak13}. But this hierarchy is not tight with the gap between classes in heararchy at least not constant. We prove almost tight hierarchy P-$k$OBDD $\subsetneq$ P-$2k$OBDD for $k = o(n/\log^3 n)$.  Our result is better than both of them. It is better than first one, because $k$ is bigger, but at the same time it is not enough tight. And it is better than second one because proper inclusion of classes is proven if $k$ is $2$ times increased. Additionally, we prove almost tight hierarchy for $k-OBDD$ of superpolynomial and subexponential size. These hierarchies improve known not tight hierarchy from \cite{K16}. Our hierarchy is almost tight (with small gap), but for little bit smaller $k$. The proof of hierarchis is based on complexity properties of Boolean function {\em Reordered Pointer Jumping}, it is modification of {\em Pointer Jumping} function from \cite{nw91}, \cite{bssw96}, is based on ideas of {\em reordering} method. For probabilistic case it is not known tight hierarchy for polynomial size only for sublinear width \cite{K16}. Additionally, for more general model Probabilistic $k$-BP  Hromkovich and Sauerhoff in 2003 \cite{hs2003} proved the tight hierarchy for $k\leq\log n/3$. We proved similar almost tight hierarchy for polynomial size bounded error probabilistic $k$-OBDD with error at most $1/3$ for $k = o(n^{1/3}/\log n)$. And almost tight hierarchies for superpolynomial and subexponential size, these results improve results from \cite{K16}. Note that, for example for nondeterministic $k$-OBDD we cannot get result better than \cite{K16}, because for constant $k$ $1$-OBDD of polynomial size and $k$-OBDD compute the same Boolean functions \cite{bhw2006}.

Structure of the paper is following. Section \ref{sec:prlmrs} contains description of models, classes and other necessary definitions. Discussion reordering method and applications for quantum OBDD located in  Section \ref{sec:deord}. The width hierarchies for quantum and probabilistic OBDDs are proved in Section \ref{sec:hierarchies}. Finally, Section \ref{sec:kobdd-hrch} contains applying reordering method and hierarchy results to deterministic and probabilistic $k$-OBDD.

\section{Preliminaries}\label{sec:prlmrs}

Ordered  read ones  Branching Programs (OBDD) are well known model for Boolean functions computation. A good source for different models of branching programs is the book by I. Wegener  \cite{Weg00}.

A branching program  over a set $X$ of $n$ Boolean variables is
a directed acyclic graph with two distinguished nodes $s$ (a source node) and $t$ (a sink node). We denote  such program $P_{s,t}$ or just $P$.   Each inner node $v$  of $P$ is associated with a variable $x\in X$. {\em Deterministic} $P$ has exactly two outgoing edges labeled $x=0$   and $x=1$ respectively for such node $v$.

The program $P$ computes the Boolean function $f(X)$ ($f:\{0,1\}^n \rightarrow \{0,1\}$) as follows: for each $\sigma\in\{0,1\}^n$ we let $f(\sigma)=1$ if and only if there exists at least one $s-t$ path (called {\em accepting} path for $\sigma$) such that all edges along this path are consistent with $\sigma$.

A branching program is {\em leveled} if the nodes can be partitioned into levels $V_1, \ldots, V_\ell$ and a level $V_{\ell+1}$ such that the nodes in $V_{\ell+1}$ are the sink nodes, nodes in each level $V_j$ with $j \le \ell$ have outgoing edges only to nodes in the next level $V_{j+1}$. For a leveled $P_{s,t}$ the source node $s$ is a node from the first level  $V_1$ of nodes  and the sink node $t$ is a node from the last level $V_{\ell+1}$.

The {\em width} $w(P)$ of a leveled branching program $P$ is the maximum
of number of nodes in  levels of $P$. $ w(P)=\max_{1\le j\le \ell}|V_j|. $ The {\em size} of branching program $P$ is a number of  nodes of
program $P$.

A leveled branching program is called {\em oblivious} if all inner
nodes of one level are labeled by the same variable.  A branching
program is called {\em read once} if each variable is tested on each
path only once. 
An oblivious leveled read once branching program is also called Ordered  Binary Decision Diagram (OBDD).
OBDD $P$ reads variables in its individual  order
$\pi=(j_1,\dots,j_n)$, $\pi(i)=j_i$, $\pi^{-1}(j)$ is position of $j$ in permutation $\pi$. We call $\pi(P)$ the order of $P$. Let us denote natural order as $id=(1,\dots,n)$. Sometimes we will use notation $id$-OBDD $P$, it means that $\pi(P)=id$. Let $width(f)=\min_{P}w(P)$ for OBDD $P$ which computes $f$ and $id\!-\!width(f)$ is the same but for $id$-OBDD.

The Branching program $P$ is called $k$-OBDD if it consists from $k$ layers, where  $i$-th ($1\le i\le k$) layer  $P^i$ of $P$  is  an OBDD. Let  $\pi_i$ be an order of $P^i$, $1\le i\le k$ and $\pi_1=\dots=\pi_k=\pi$.
 We call  order
$\pi(P)=\pi$ the order of $P$.

Let $tr_P:\{1,\dots,n\}\times\{1,\dots,w(P)\}\times\{0,1\}\to\{1,\dots,w(P)\}$ be transition function of OBDD $P$ on level $i$.
OBDD $P$ is called {\em commutative} if for any permutation $\pi'$ we can construct OBDD $P'$ by just reordering transition functions and $P'$ still  computes the same function. Formally, it means $tr_{P'}(i,s,x_{\pi'(i)})=tr_{P}(\pi^{-1}(\pi'(i)),s, x_{\pi'(i)})$, for $\pi$ is order of $P$, $i\in \{1,\dots,n\}$, $s\in \{1,\dots,w(P)\}$. $k$-OBDD $P$ is commutative if each layer is commutative OBDD.

Nondeterministic OBDD (NOBDD) is nondeterministic counterpart of OBDD.
Probabilistic OBDD (POBDD) can have more than two edges for node, and choose one of them using probabilistic mechanism. POBDD $P$ computes Boolean function $f$ with bounded error $0.5-\varepsilon$ if probability of right answer is at least $0.5+\varepsilon$.


Let us discuss a definition of quantum OBDD (QOBDD). It is given in different terms, but you can see that it is equivalent. You can see \cite{agkmp2005}, \cite{agk01} for more details.

For a given $ n>0 $, a quantum OBDD $ P$ of width $w$, defined on $ \{0,1\}^n $, is a 4-tuple
$
	P=(T,\ket{\psi}_0,Accept,\pi),
$
where
\begin{itemize}
	\item $ T = \{ T_j : 1 \leq j \leq n \mbox{ and } T_j = (G_j^0,G_j^1)  \} $ are  ordered pairs of (left) unitary matrices representing the transitions is applied at the $j$-th step, where $ G_j^0 $ or $ G_j^1 $, determined by the corresponding input bit, is applied.
	\item $\ket{\psi}_0$ is initial vector from $ w $-dimensional Hilbert space over field of complex numbers.  $ \ket{\psi}_0=\ket{q_0}$ where $ q_0 $ corresponds to the initial node.
	\item $ Accept \subset \{1,\ldots,w\} $ is accepting nodes.
\item $ \pi $ is a permutation of $ \{1,\ldots,n\} $ defining the order of testing the input bits.
\end{itemize}

  For any given input $ \sigma \in  \{0,1\}^n $, the computation of $P$ on $\sigma$ can be traced by  a vector from $ w$-dimensional Hilbert space over field of complex numbers. The initial one is $ \ket{\psi}_0$. In each step $j$, $1 \leq j \leq n$, the input bit $ x_{\pi(j)} $ is tested and then the corresponding unitary operator is applied:
$
	\ket{\psi}_j = G_j^{x_{\pi(j)}} (\ket{\psi}_{j-1}),
$ 
where $ \ket{\psi}_{j-1} $ and $ \ket{\psi}_j $ represent the state of the system after the $ (j-1)$-th  and $ j$-th steps, respectively, where $ 1 \leq j \leq n $.

In the end of computation program $P$ measure qubits. The accepting (return $1$) probability $Pr_{accept}(\sigma)$ of $ P_n $ on input $ \sigma $ is 
$
	Pr_{accept}(\nu)=\sum_{i \in Accept} v^2_i.
$, for $ \ket{\psi}_n=(v_1,\dots,v_w)$.
We say that a function $f$ is computed by $ P$ with bounded error if there exists an $ \varepsilon \in (0,\frac{1}{2}] $ such that $ P$ accepts all inputs from $ f^{-1}(1) $ with a probability at least $ \frac{1}{2}+\varepsilon $ and $ P_n $ accepts all inputs from $ f^{-1}(0) $ with a probability at most $ \frac{1}{2}-\varepsilon $.  We can say that error of answer is $\frac{1}{2}-\varepsilon$.

\section{Reordering Method and Exponential Gap Between Quantum and Classical OBDD}\label{sec:deord}
Let us introduce some helpful definitions.
Let $\theta=(\{x_{j_1},\dots, x_{j_u}\}, \{x_{i_1},\dots, x_{i_{n-u}}\})=(X_A,X_B)$
 be a partition of set $X$ into two parts. Below we will use equivalent notations $f(X)$ and $f(X_A, X_B)$.
Let  $f|_\rho$ be a subfunction of $f$, where  $\rho$ is a mapping $\rho:X_A \to \{0,1\}^{|X_A|}$.
Function $f|_\rho$ is obtained from $f$ by applying $\rho$, so if $\rho:X_A\to \nu$, then $f|_\rho(X_B)=f(\nu,X_B)$. Let $N^\theta(f)$  be number of different subfunctions with respect to partition $\theta$. 
Let $\Pi(n)$ be the set of all permutations of $\{1,\dots,n\}$.
We say, that  partition $\theta$ agrees with permutation
$\pi=(j_1,\dots, j_n)\in \Pi(n)$, if for some $u$, $1<u<n$ the
following is right: $\theta=(\{x_{j_1},\dots,
x_{j_u}\},\{x_{j_{u+1}},\dots, x_{j_n}\})$. We denote $\Theta(\pi)$
a set of all partitions which agrees with $\pi$. 
Let $ N^\pi(f)=  \max_{\theta\in \Theta(\pi)} N^\theta(f),
N(f)=\min_{\pi\in \Pi(n)}N^\pi(f)$.


It is known that the difference between quantum and deterministic OBDD complexity is at most exponential \cite{agkmp2005}. But one of the main issues in proof of complexity of OBDD is different orders of input variables. We suggest a method, called ``reordering'', which allows to construct partial function $f'$ from Boolean function $f$  such that $N^{id}(f)=d(n), N(f')\geq d(q),  n=q(\lceil\log q\rceil+1)$. Note that $N(f')=width(f')$ and $N^{id}(f)=id\!-\!width(f)$, due to \cite{Weg00}. At the same time, if commutative QOBDD $P$ of width $g(n)$ computes $f$, then  we can construct QOBDD $P'$ of width $g(q) \cdot q$ which computes $f'$. If $g(n)=O(n)$ and $d(n)=O(2^n)$, then we can say that $d(q/\lceil\log q +1\rceil)$ is almost exponential great than $g(q) \cdot q$. And total boolean function $f''$ with same properties can be built using  result of computation of $P'$ for unspecified inputs.
 And for some functions we can give explicit definition of such total {\em reordered} function.
 
{\bf Reordering Method.}
Let us shuffle input bits for solving ``order issues''. It means that order of {\em value} bits is determined by input. Let us consider input $X=(x_1,\dots,x_n)$, among the variables we have $q$ {\em value} bits $Z=\{z_1, \ldots ,z_q\}$, where $q$ is such that $n=q(\lceil\log q\rceil+1)$. And any {\em value} bit has $\lceil\log q\rceil$ bits as {\em address}, that is binary representation of number of real position of {\em value} bit in input. We call this process as {\em reordering} of input or {\em reordering} of Boolean function $f(X)$. Now from $f(X)$ we obtain a new partial Boolean function $f'(X)$ on reordered input, such that any {\em value} bit has unique address and all addresses from $\{1,\dots,q\}$ are occurred in input. In a case of {\em xor-reordering} address of {\em value} bit can be obtain as parity of current and previous {\em address} bits.  

Let us formally describe  a partial function $f'(X)$ 
:
\begin{itemize}
\item $X$ consists of $q$ blocks, for $n=q(\lceil\log q\rceil+1)$ or $q= O(n/\log{n})$.

\item Block $i$ consists of $p=\lceil\log{q}\rceil$ {\em address} bits $y^i_1,\dots,y^i_{p}$ and one {\em value} bit $z^i$. Formally, $(x_{(i-1)(p+1)+1},\dots,x_{i(p+1)})=(y^i_1,\dots,y^i_{p},z^i)$.

\item Function $Adr:\{0,1\}^n\times\{ 1, \ldots ,q\} \to \{ 1, \ldots ,q\}$, $Adr(X,i)$ is the address of $i$-th {\em value} bit. Let $bin(y_1,\dots,y_p)$ is a number, which binary representation is  $(y_1,\dots,y_{p})$, then in a case of {\em reordering} $Adr(X,i)=bin(y^i_1,\dots,y^i_{p})+1$. If we consider {\em xor-reordering} then $Adr(X,i)=Adr'(X,i)+1$, $Adr'(X,i)=Adr'(X,i-1)\oplus bin(y^i_1,\dots,y^i_{p})$ for $i\geq 1$ and $Adr'(X,0)=0$. Here when we apply parity function to integers, we mean parity of their bits in binary representation.
\item We consider only such inputs $\sigma\in\{0,1\}^n$ that addresses of the blocks are different and all addresses are occurred. Formally, $\{1,\dots,q\}=\{Adr(\sigma,1),\dots,Adr(\sigma,q)\}$.  
\item Let a permutation $\pi=(Adr(\sigma,1),\dots,Adr(\sigma,q))$, and $\gamma$ is string of {\em value} bits of $\sigma$ then $f'(\sigma)=f(\gamma_{\pi^{-1}(1)},\dots, \gamma_{\pi^{-1}(q)})$
\end{itemize}

\begin{figure}[tbh]
\begin{center}
\includegraphics[width=8cm]{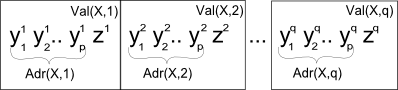}
\end{center}
\caption{Input. Blocks of {\em address} and {\em value} bits.}
\label{fig:address-value}
\end{figure}

\begin{theorem} \label{th:d-obdd}
Let Boolean function $f$ over $X=(x_1, \cdots, x_n)$, such that $N^{id}(f)\geq d(n)$. Then partial Boolean function $f'$, reordered or xor-reordered version of $f$, such that $N(f')\geq d(q)$, where  $n=q(\lceil\log q\rceil+1)$.
\end{theorem}
\Beginproof
Let us consider function $f'(X)$ and any order $\pi=(j_1,\dots,j_n)$ and $\pi'=(i_1,\dots,i_q)$ is the order of {\em value} bits according to order $\pi$. Let $\Sigma$ be the set of inputs with natural order of blocks ({\em value} bits) with respect to $\pi$, that is $\Sigma=\{\sigma\in\{0,1\}^n: Adr(\sigma,i_r)=r,$ for $1\leq r \leq q\}$. 
Let partition $\theta\in \Theta((1,\dots,q)), \theta=(\{x_1,\dots,x_u\}, \{x_{u+1},\dots,x_q\})$ be such that $N^\theta(f)=N^{id}(f)$. And let partition $\theta'=(X_A,X_B)=(\{x_1,\dots,x_{u'}\}, \{x_{u'+1},\dots,x_n\})$ , for $\theta'\in \Theta(\pi)$, be such that exactly $u$ {\em value} bits belongs to $X_A$ and others to $X_B$.
Let $\Gamma=\{\gamma\in\{0,1\}^u:$ for different $\gamma,\gamma'$ and corresponding subfunctions holds $f|_\rho\neq f|_{\rho'}\}$. And $\Xi=\{\xi\in\{0,1\}^{u'}:$ there are $\nu\in\{0,1\}^{n-u'}$ such that $(\xi,\nu)\in \Sigma$ and string of {\em value} bits of $\xi$ belongs to $\Gamma\}$.
It is easy to see that $|\Gamma|=|\Xi|$ and each $\xi\in\Xi$ produce own subfunction of function $f'$. Therefore $N^{\theta'}(f')\geq N^{\theta}(f)$.  Due to definition, $N^{\pi}(f')\geq N^{\theta'}(f')\geq N^{id}(f)$. It is right for any order, hence $N(f')\geq N^{id}(f)$.
\Endproof

\begin{theorem}\label{th:d-qobdd}
Let commutative QOBDD $P$ of width $g=g(n)$ computes a Boolean function $f(X)$ over  $X=(x_1, \ldots, x_n)$. Then there is $id$-QOBDD $P'$ of width $g(q) \cdot q$ which computes partial Boolean function $f'$, xor-reordered version of $f$, where $q$ is such that $n=q(\lceil\log q\rceil+1)$. 
\end{theorem}

\Beginproof
Because of $P$ is commutative, we can consider $id$-QOBDD $P_{id}$ of the same width $g(n)$ for function $f$.
For description of computation of $P'$ we use quantum register $\ket\psi = \ket {\psi_1 \psi_2 \ldots \psi_t}$, where $t=\lceil \log{g} \rceil$ and $g \times g$ matrices for unitary operators $(G_i^0, G_i^1)$, $i \in \{1,2 \ldots q\}$.

Then we consider partial function $f'(X)$ described above. In this case we have $q$ {\em value} bits and $p=\lceil \log{q} \rceil$ {\em address} bits for any {\em value} bit. Let us construct QOBDD $P'$ for computing $f'$.
%

Program $P'$ has quantum register of $\lceil \log{g} \rceil + \lceil \log{q} \rceil$ qubits, having $g \cdot q$ states. Let us denote it as $\ket\phi = \ket {\phi_1 \phi_2 \ldots \phi_p \psi_1 \psi_2 \ldots \psi_t}$, where $t=\lceil \log{g} \rceil$, $p=\lceil \log{q} \rceil$. 

Part of register $\ket\phi$ consisting of  $\ket{\psi_1 \psi_2 \ldots \psi_t}$ qubits (we note it as a {\em computing} part) is modified on reading {\em value} bit. In other hand, we added qubits $\ket{\phi_1 \phi_2 \ldots \phi_p }$ (let this part be an {\em address} part) to determine address of {\em value} bit. And superposition of the states of these two parts will give us a right computation of function. 
Program $P'$ consists of $q$ parts, because input contains $q$ blocks, for $n=(\lceil\log q\rceil+1)q$ or $q= O( n/\log{n})$. For any $i \in \{1,\dots,q\}$ block $i$ handles {\em value} bit $z^i$.

Informally, when $P'$ processes a block, it stores address in {\em address} part by applying parity function. After that some modifications are produced on the {\em computation} part, with respect to {\em value} bit. 

Let us describe $i$-th block of levels formally, for $i\in\{1,\dots,q\}$. In the first $\lceil \log q \rceil$ levels the program computes address $Adr'(X,i)$, it reads bits one by one, and for bit $y_j^i$  we use unitary operator $U^{y^i_j}_j$ on the {\em address} part of register $\ket\phi$, for $j \in \{1,2 \ldots, p\}$ (see Picture \ref{fig:address-value}). $U^{y^i_j}_j=I \otimes I \otimes \ldots \otimes I \otimes A^{y^i_j} \otimes I \ldots \otimes I $, where $A^{0}=I$ and $A^{1}=NOT$, $I$ and $NOT$ are $2\times 2$ matrices, such that $I$ is diagonal $1$-matrix and $NOT$ is anti-diagonal $1$-matrix. And we do not modify {computation} part. 

After these operations {\em address} part of register in binary notation equals to the address $Adr'(X,i)$. In the vector of the states all elements are equals to zero except elements of block where address part of qubits corresponds to $Adr(X,i)$. 

After reading $z^i$ we transform system  $\ket\phi$ by unitary $(g\cdot q \times g\cdot q)$-matrix $D^{z^i}$.

$D^0 = \begin{pmatrix}
G_1^0 & 0 & \cdots & 0 \\
0 & G_2^0 & \cdots & 0 \\         
\vdots & \vdots & \ddots & \vdots \\
0 & 0 & \cdots & G_q^0
\end{pmatrix}$ and
$D^1 = \begin{pmatrix}
G_1^1 & 0 & \cdots & 0 \\
0 & G_2^1 & \cdots & 0 \\         
\vdots & \vdots & \ddots & \vdots \\
0 & 0 & \cdots & G_q^1
\end{pmatrix}$,

where matrices $\{(G_i^0,G_i^1), 1\leq i \leq q\}$ are unitary matrices transforming quantum system in  $id$-QOBDD $P_{id}$. 

Because of size of register, QOBDD $P'$  has width $g(q) \cdot q$.  Let us prove that $P'$ computes $f'$.

Let us consider an input $\sigma\in\{0,1\}^n$.  Let a permutation $\pi = (j_1, \ldots, j_q)=(Adr(\sigma,1),\dots, Adr(\sigma,q))$ be an order of {\em value} variables with respect to input $\sigma$.   

Due to $id$-QOBDD $P_{id}$ is commutative, we can reorder unitary operators $\{(G_i^0,G_i^1), 1\leq i \leq q\}$ according to order $\pi$ and get a QOBDD $P_{\pi}$ computing $f$ as well.

It is easy to see that $P'$ exactly emulates computation of $P_{\pi}$, therefore $P'$ on $\sigma$ gives us the same result as $P_{\pi}$ on corresponding {\em value} bits. So, by definition of $f'$ we have $P'$ computes $f'$.
\Endproof
\begin{CR}
Let commutative $k$-QOBDD $P$ of width $g=g(n)$ computes a Boolean function $f(X)$ over  $X=(x_1, \ldots, x_n)$. Then there is $k$-QOBDD $P'$ of width $g(q) \cdot q$ which computes partial Boolean function $f'$, xor-reordered version of $f$, where $q$ is such that $n=q(\lceil\log q\rceil+1)$.
\end{CR}

It is proved exactly by the same way as Theorem \ref{th:d-qobdd}.

\begin{theorem}\label{th:deord-classic}
If for some Boolean function $f:\{0,1\}^n\to \{0,1\}$ there are commutative $k$-OBDD $P_1$, $k$-NOBDD $P_2$ and $k$-POBDD $P_3$ that computes $f$ and width of $P_i$ is $d_i$ for $i\in\{1,2,3\}$. 
Then there are $k$-OBDDs $D_1,D_4$, $k$-NOBDDs $D_2,D_5$  and $k$-POBDDs $D_3,D_6$ such that width of $D_i$ is $d_i(q)\cdot q$ for $i\in\{1,2,3\}$ and $d_{i-3}(q)\cdot q$ for $i\in\{4,5,6\}$, and $D_i$ computes  $f'$, reordered version of $f$, and $D_j$ computes $f''$, xor-reordered version of $f$, for $i\in\{1,2,3\}$, $j\in\{4,5,6\}$, $n=q(\lceil\log q\rceil+1)$, $f'$ and $f''$ are partial Boolean function.
\end{theorem}
\Beginproof 
Let $P_1$ be commutative deterministic $k$-OBDD of width $d_1(n)$ which computes Boolean function $f$. Let Boolean function $f'$ is  reordered $f$ and $f''$ is xor-reordered $f$, note that $f'$ and $f''$ are partial Boolean functions.

We want to construct deterministic $k$-OBDDs $D_1$ and $D_4$ of width $q\cdot d_1(q)$, for $n=q(\lceil \log_2  q \rceil+1)$. $D_1$ is for reordering case and $D_4$ is for xor-reordering one. $D_1$ and $D_4$  read variables in natural order. $D_1$ and $D_4$ have $q\cdot d_1(q)$ nodes on level, each of them corresponds to pair $(i,s)$, where $i\in\{1,\dots,q\}$, $s\in\{1,\dots, d(q)\}$. Let us describe computation on  block $j$. 

\begin{itemize}
\item A case of reordering and program $D_1$.  In the begin of the block $D_1$ situated in one of the nodes $(1,s)$. After reading first $\lceil\log q\rceil $ bits of the block $D'$ just store a number $Adr(X,j)=a$ in states and program reaches node corresponding to $(a,s)$. Then if transition function of $P$ is such that $s'=tr_D(\pi^{-1}(a),s, z^j)$ then $D'$ reaches $(1,s')$ .

\item A case of xor-reordering  and program $D_4$.  In the begin of the block $D_4$ is situated in one of the nodes $(b,s)$. After reading first $\lceil\log q\rceil $ bits of the block $D_4$ just computes parity of $b-1$ and {\em address} bits, so computes a number $Adr'(X,j)=a'$ and program reaches node, which is correspond to $(a,s)$, for $a=a'+1$.
\end{itemize}

In the case when all addresses are different, $D_1$ and $D_4$ just emulate work of $D_{\pi}$ which is constructed from $D$ by permutation of transition function with respect to order $(Adr(X,1),\dots,Adr(X,q))$. By the definition of commutative $k$-OBDD the $D_{\pi}$ computes the same function $f$. Therefore $D_1$ and $D_4$ also return the same result. And by the definition of functions $f'$ and $f''$ programs $D_1$ computes $f'$ and $D_4$ computes $f''$.

We can construct nondeterministic $k$-OBDDs $D_2, D_5$ and probabilistic $k$-OBDDs $D_3, D_6$ by the similar way.
\Endproof

\begin{CR}\label{cr:total-reordering}
Let Boolean function $f$ over $X=(x_1, \cdots, x_n)$, such that $N^{id}(f)\geq d(n)$ and commutative k-QOBDD $P$, $k$-OBDD $D$, $k$-NOBDD $H$ and $k$-POBDD $R$ of width $g(n), d(n), h(n)$ and $w(n)$, respectively , computes $f$. Then there are total Boolean functions $f^{(i)}$, total xor-reordered version of $f$, for $i\in\{1,\dots,4\}$ and $f^{(j)}$, total reordered version of $f$, for $j\in\{5,\dots,7\}$,  such that $N(f^{(i)}), N(f^{(j)}),\geq d(q)$, where  $n=q(\lceil\log q\rceil+1)$. And there are $k$-QOBDD $P'$ of width $g(q) \cdot q$ which computes $f^{(1)}$ and $k$-OBDD $D'$, $k$-NOBDD $H'$ and $k$-POBDD $R'$ of  width $d(q)\cdot q, h(q)\cdot q$ and $w(q)\cdot q$, respectively, such that $D'$ computes computing $f^{(2)},f^{(5)}$, $H'$ computes computing $f^{(3)},f^{(6)}$, $R'$ computes computing $f^{(4)},f^{(7)}$. 
\end{CR}
\Beginproof
Let partial Boolean function $f'$ be xor-reordered version of $f$. Due to Theorems \ref{th:d-qobdd} and \ref{th:d-obdd}, $N(f')\geq d(q)$, $id$-QOBDD $P'$ of width $g(q) \cdot q$ computes $f'$. Let total function $f^{(1)}$ be such that $f^{(1)}(\sigma)=f'(\sigma)$ for input $\sigma$ allowed for $f'$. And for input $\sigma'$, not allowed for $f'$, $f^{(1)}(\sigma)$ equals to result of $P'$.
It is easy to see that $N(f^{(1)})\geq N(f')$. Similar prove for $f^{(i)}$ and $f^{(j)}$.
\Endproof

{\bf Exponential Gap Between Quantum and Classical OBDDs.}
Let us apply reordering method to {\em Equality function} $EQ_n:\{0,1\}^n\to \{0,1\}$. $EQ_n(\sigma)=1$, iff $(\sigma_1,\sigma_2 \ldots \sigma_{n/2})=(\sigma_{n/2+1}, \sigma_{n/2+2} \ldots \sigma_{n})$.

From \cite{akv2008} we know, that there is commutative $id$-QOBDD $P$ of width $O(n)$, which computes this function. After xor-reordering we get partial function $EQ_n'(X)$ computed by $QOBDD$  of width $O(q) \cdot q=O(q^2)$, where $q= O( n/\log{n})$.

It is known that $N^{id}(EQ_n)=2^{n/2}$. Due to Theorem \ref{th:d-obdd} we have $N(EQ'_n)\geq 2^{q/2}$, therefore deterministic OBDD has width at least $2^{q/2}$.

So we have following Theorem for $EQ'_n$:

\begin{theorem}
Let partial Boolean function $EQ'_n$  is xor-reordered $EQ_n$. Then there is quantum OBDD $P$ of width $O(n^2/\log^2 n)$  which computes $EQ'_n$ and any deterministic OBDD $D$  which computes $EQ'_n$ has width  $2^{\Omega(n/\log n)}$.
\end{theorem}

Let us consider {\em Reordered Equality function} $REQ_n:\{0,1\}^n\to \{0,1\}$. This is total version of $EQ'_n$ and on inputs which is not allowed for $EQ'_n$ the result of function is exactly result of $QOBDD$ $P'$ which was contracted for $EQ'_n$ by the method from the proof of Theorem \ref{th:d-qobdd}. Due to fingerprinting algorithm for $EQ_n$ from \cite{akv2008}, we can see that  $REQ_n(\sigma)=1$ iff $\sum_{i=1}^{q/2}2^{Adr'(\sigma,i)}Val(\sigma,i)=\sum_{i=q/2+1}^{q}2^{Adr'(\sigma,i)}Val(\sigma,i)$.
We can prove the following lemma for this function: 

\begin{lemma}\label{lm:deq-dlower} $N(REQ_n)\geq 2^{n/(2\lceil \log_2 n +1\rceil)}$ .
\end{lemma}
It means that any determenistic OBDD $P$ of width $w$ computing $REQ_n$ is such that $w\geq 2^{n/(2\lceil \log_2 n + 1\rceil)}$.  
Therefore function $REQ$ such that:
\begin{theorem} \label{th:deq}
There is quantum OBDD $P$ of width $O(n^2/\log^2 n)$  which computes total Boolean function $REQ_n$ and any deterministic OBDD $D$  which computes $REQ$ has width  $2^{\Omega(n/\log n)}$.
\end{theorem}
\Beginproof
By the definition of the function we can construct $QOBDD$ $P$ and Lemma \ref{lm:deq-dlower} shows a bound for deterministic case.
\Endproof

So, $REQ_n$ is explicit function which shows the following distance between quantum and deterministic ODDD complexity:  $O(n^2/\log^2 n)$ and  $2^{\Omega(n/\log n)}$.

\section{Hierarchy for Probabilistic and Quantum OBDDs}\label{sec:hierarchies}
Let us consider classes {\bf BPOBDD}$_d$ and {\bf BQOBDD}$_d$ of Boolean functions  that will be computed by probabilistic and quantum OBDDs with bounded error of width $d$, respectively. We want to prove a width hierarchy for these classes. 

{\bf Hierarchy for Probabilistic OBDDs.} 
Before proof of hierarchy let us consider the Boolean function $WS_n(X)$ due to Savick{\`y} and {\v{Z}}{\'a}k \cite{sz2000}. For a positive integer $n$ and $X = (x_1, \dots , x_n) \in \{0, 1\}^n$, let $p(n)$ be the smallest
prime larger than $n$ and let $s_n(X) = \left(\sum_{i=1}^{n} i\cdot x_i\right)
\mbox{ }mod\mbox{ }p(n)$. Define the weighted
sum function by $WS_n(x) = x_{s_n(X)}$.
For this function it is known that for every $n$ large enough it holds that any bounded error probabilistic OBDD $P$ which computes $WS_n(X)$ has size no less than $2^{\Omega(n)}$.  
Let us  modify Boolean function $WS_n(X)$ using pending bits. We will denote it $WS^b_n(X)$. For a positive integers $n$ and $b$, $b\leq n/3$ and $X = (x_1, \dots , x_n) \in \{0, 1\}^n$, let $p(b)$ be the smallest
prime larger than $b$, $s_b(X) = (\sum_{i=1}^{b} i\cdot x_i)
\mbox{ }mod\mbox{ }p(b)$. Define the weighted
sum function by $WS_n^b(x) = x_{s_b(X)}$.
We can prove the following lemma by the way as in \cite{sz2000}.

\begin{lemma}\label{lm:ws}
For large enough $n$ and $const=o(b)$,  any bounded error probabilistic OBDD $P$ computing $WS_n^b(X)$ has width no less than $2^{\Omega(b)}$. There is  bounded error probabilistic OBDD $P$ of width $2^b$ which computes $WS_n^b(X)$.  
\end{lemma}

The second claim of the Lemma follows form the fact that any Boolean function over $X  \in \{0, 1\}^n$ can be computed by deterministic OBDD of width $2^n$, just by building full binary tree.


Let us prove hierarchy for {\bf BPOBDD}$_d$ classes using these properties of Boolean function $WS_n^b(X)$.

\begin{theorem}\label{th:prob-hi}
For integer $d=o(2^n), const = o(d)$, the following statement holds: 

{\bf BPOBDD}$_{d^{1/\delta}}\subsetneq${\bf BPOBDD}$_d$, for  $const=o(\delta)$. 
\end{theorem}
\Beginproof
It is easy to see that {\bf BPOBDD}$_{d^{1/\delta}}\subseteq${\bf BPOBDD}$_d$. Let us prove inequality of these classes.
Due to Lemma \ref{lm:ws}, Boolean function $WS_n^{\log d}\in${\bf BPOBDD}$_d$, at the same time for any bounded error probabilistic OBDD $P$ we have $w(P)=2^{\Omega(\log d)}>2^{(\log d)/\delta}=d^{1/\delta}$. Therefore $WS_n^{\log d}\not\in${\bf BPOBDD}$_{d^{1/\delta}}$.
\Endproof

{\bf Hierarchy for Quantum OBDDs. }
Let us  modify Boolean function $REQ_n(X)$ using pending bits as for $WS_n^b(X)$. We will denote it $REQ^b_n(X)$. Also let us consider complexity property of $MOD_p$ function (number of $1$s by modulo $p$ is $0$). And Boolean function $MSW_n^b(X)$, it is similar modification of $MSW_n(X)$ function \cite{s06}  using pending bits. $MSW_n^b(X)=x_z\oplus x_{r+n/2}$, where $z=s_{b/2}(x_1,\dots,x_{b/2}), r=s_{b/2}(x_{b/2+1},\dots,x_{b})$, if $r=z$ and $MSW_n^b(X)=0$ otherwise. Complexity properties of functions are described in the following lemma.




\begin{lemma}\label{lm:deqb-modp-mwsb}
Claim 1.
Any bounded error quantum OBDD $P$ which computes $REQ_n^b(X)$ has width at least $\lfloor b/\lceil \log b + 1\rceil \rfloor$, for $\lfloor b/\lceil \log b + 1\rceil \rfloor\geq 1$. 
There is  bounded error quantum OBDD $P$ of width $b^2$ which computes $REQ_n^b(X)$.  

Claim 2.
Any bounded error quantum OBDD $P$ which computes $MOD_p(X)$ has width no less than $\lfloor \log p \rfloor$, for $2\leq p \leq n$. 
There is  bounded error quantum OBDD $P$ of width $O(\log p)$ which computes $MOD_p(X)$.  

Claim 3.
Any bounded error quantum OBDD $P$ which computes $MSW_n^b(X)$ has width no less than $2^{\Omega(b)}$, for $const=o(b)$. 
There is  bounded error quantum OBDD $P$ of width $2^b$ which computes $MSW_n^b(X)$.  
\end{lemma}

A proof of Claim 1 is similar to Theorem \ref{th:deq}, a proof of Claim 2 is presented in \cite{agkmp2005}, \cite{av2008} and a proof of Claim 3 is based on result from \cite{s06}.

  Let us prove hierarchies for {\bf BQOBDD}$_d$ classes using presented above lemma.

\begin{theorem} \label{th:quntum-hi} For a integer $d$ following statements are right:

{\bf BQOBDD}$_{d/\delta^2}\subsetneq${\bf BPOBDD}$_{d^2}$, for $d<\log n, d>2$,   $const=o(\delta)$.

{\bf BQOBDD}$_{d/\log^2_2 d}\subsetneq${\bf BPOBDD}$_{d^2}$, for $d<n\, d>2$.

{\bf BQOBDD}$_{d^{1/\delta}}\subsetneq${\bf BQOBDD}$_d$, for $d=o(2^n), const = o(d)$, $const=o(\delta)$.
\end{theorem}
A proof is based on Lemma \ref{lm:deqb-modp-mwsb}. 

\section{Extension of Hierarchy for Deterministic and Probabilistic $k$-OBDD}\label{sec:kobdd-hrch}

Let us apply the reordering method to $k$-OBDD model. We will prove almost tight hierarchy for Deterministic and Probabilistic $k$-OBDDs using complexity properties of {\em Pointer jumping} function ($PJ$) \cite{nw91}, \cite{bssw96}. These hierarchies are  extention of excisting ones. At first, let us present version of function which works with integer numbers.

 Let $V_A,V_B$ be two disjoint sets (of vertices) with $|V_A| = |V_B| = m$
and $V = V_A \cup V_B$ . Let $F_A = \{f_A : V_A \to V_B\}$, $F_B = \{f_B : V_B \to V_A\}$ and $f = (f_A, f_B):V \to V$ defined by $f(v) = f_A(v)$, if $v\in V_A$ and $f= f_B(v)$, $v\in V_B$. For each $k \geq 0$ define $f^{(k)}(v)$ by $f^{(0)}(v) = v$ , $f^{(k+1)}(v) = f(f^{(k)}(v))$ .
Let $v_0\in V_A$. The function we will be interested in computing is $g_{k,m} : F_A \times F_B \to V$ defined by $g_{k,m}(f_A, f_B) = f^{(k)}(v_0)$. Boolean function $PJ_{t,n}:\{0,1\}^n\to\{0,1\}$ is boolean version of $g_{k,m}$, where we encode $f_A$ in a binary string using $m\log m$ bits and do it with  $f_B$ as well. The result of function is parity of binary representation of result vertex.

Let us apply reordering method to $PJ_{k,n}$ function. $RPJ_{k,n}$ is total version of reordered $PJ_{k,n}$.
Formally:
Boolean function $RPJ_{k,n}:\{0,1\}^n\to\{0,1\}$ is following. Let us separate whole input $X=(x_1,\dots,x_n)$ to $b$ blocks, such that $b\lceil\log_2b+1\rceil = n$, therefore $b=O(n/\log n)$. 
And let $Adr(X,i)$ be integer, which binary representation is first $\lceil \log_2 b \rceil$ bits of $i$-th block and $Val(X,i)$ be a value of bit number $\lceil \log_2 b+1 \rceil$ of block $i$, for $i\in\{0,\dots,b-1\}$. Let $a$ be such that $b=2a\lceil\log_2 a\rceil$ and $V_A=\{0,\dots,a-1\}$, $V_B=\{a,\dots,2a-1\}$. 

Let function $BV:\{0,1\}^n\times\{0,\dots, 2a-1\}\to \{0,\dots, a-1\}$ be the following:
\[BV(X,v) = \sum_{i:(v-1)\lceil\log_2 b \rceil< Adr(X,i)\leq v\lceil\log_2 b \rceil}2^{Adr(X,i) - (v-1)\lceil\log_2 b \rceil}\cdot Val(X,i)\mbox{ }(mod\mbox{ }a)\]

Then $f_A(v)=BV(X,v)+a$, $f_B(v)=BV(X,v)$.

Let $r=g_{k,a}(f_A,f_B)$, then 
\[RPJ_{k,n}(X)=\bigoplus_{i:(r-1)\lceil\log_2 b \rceil < Adr(X,i)\leq r\lceil\log_2 b \rceil}Val(X,i)\]. 

Let us prove lower bound for this function:

\begin{lemma}\label{lm:dpj}

Claim 1. The functions $RPJ_{2k-1,n}$ can be computed by $2k$-OBDD of size $O(n^3)$.

Claim 2. Each $k$-OBDD for $RPJ_{2k-1,n}$, has size $2^{\Omega(n/(k\log n)-\log (n/\log n))}$. Each $k$-POBDD for $RPJ_{2k-1,n}$ which computed with bounded error at least $1/3$, has size $2^{\Omega( n/(k^3\log n)  - \log(n/\log n))}$.

\end{lemma}
A proof of lower bound is based on communication complexity properties of the function $PJ_{k,n}$ from \cite{nw91}. And a proof of upper bound is based on Theorem \ref{th:deord-classic} and Corollary \ref{cr:total-reordering}.

Using this lemma we extend  hierarchy for following classes: P-$k$OBDD,  BP$_{1/3}$-$k$OBDD, SUPERPOLY-$k$OBDD,  BSUPERPOLY$_{1/3}$-$k$OBDD, SUBEXP$_{\alpha}$-$k$OBDD and BSUBEXP$_{\alpha,1/3}$-$k$OBDD. These are classes of Boolean functions computed by following models: 
\begin{itemize}
\item P-$k$OBDD and  BP$_{\delta}$-$k$OBDD are for polynomial size $k$-OBDD, the first one is for deterministic case and the second one is for bounded error probabilistic $k$-OBDD with error at least $\delta$.
\item SUPERPOLY-$k$OBDD and  BSUPERPOLY$_{1/3}$-$k$OBDD are similar classes for superpolynomial size models
\item SUBEXP$_{\alpha}$-$k$OBDD and BSUBEXP$_{\alpha,1/3}$-$k$OBDD are similar classes for size at most $2^{O(n^\alpha)}$, for $0<\alpha<1$.
\end{itemize}
\begin{theorem}\label{thm:hierarchy}
Claim 1.
P-$k$OBDD $\subsetneq$ P-$2k$OBDD , for $k = o(n/\log^3 n)$. BP$_{1/3}$-$k$OBDD $\subsetneq$ BP$_{1/3}$-$2k$OBDD, for $k = o(n^{1/3}/\log n)$.

Claim 2.
SUPERPOLY-$k$OBDD $\subsetneq$ SUPERPOLY-$2k$OBDD , for $k = o(n^{1-\delta})$, $\delta>0$. BSUPERPOLY$_{1/3}$-$k$OBDD $\subsetneq$ BSUPERPOLY$_{1/3}$-$2k$OBDD, for $k = o(n^{1/3-\delta})$, $\delta>0$.

Claim 3.
SUBEXP$_{\alpha}$-$k$OBDD $\subsetneq$ SUBEXP$_{\alpha}$-$2k$OBDD , for $k = o(n^{1-\delta})$,  $1>\delta>\alpha+\varepsilon$, $\varepsilon>0$. BSUBEXP$_{\alpha,1/3}$-$k$OBDD $\subsetneq$ BSUBEXP$_{\alpha,1/3}$-$2k$OBDD , for $k = o(n^{1/3-\delta/3})$, $1/3>\delta>\alpha+\varepsilon$, $\varepsilon>0$.
\end{theorem}
A proof is based on Lemma \ref{lm:dpj}. 

\noindent
\textbf{Acknowledgements.}
We thank Alexander Vasiliev and Aida Gainutdinova from Kazan Federal University and Andris Ambainis from University of Latvia for their helpful comments and discussions. 


\bibliographystyle{alpha}
\bibliography{tcs}

\newpage
\appendix

\section{Proof of Corollary \ref{cr:total-reordering}}\label{apx:total-reordering}

Let partial Boolean functions $f'$ be xor-reordered version of $f$, $f''$ be reordered version of $f$. Due to Theorems \ref{th:d-qobdd} and \ref{th:d-obdd}, $N(f')\geq d(q)$, $id$-QOBDD $P'$ of width $g(q) \cdot q$ computes $f'$. Let total function $f^{(1)}$ be such that $f^{(1)}(\sigma)=f'(\sigma)$ for input $\sigma$ allowed for $f'$. 
And for input $\sigma'$, not allowed for $f'$, $f(\sigma)$ is result of $P'$. Functions $f^{(2)}$ and $f^{(5)}$ are similar, but with respect to $D'$ and $D''$ computing $f'$ and $f''$, respectively. Functions $f^{(3)}$ and $f^{(6)}$ are similar, but with respect to $H'$ and $H''$ computing $f'$ and $f''$, respectively. Functions $f^{(4)}$ and $f^{(7)}$ are similar, but with respect to $R'$ and $R''$ computing $f'$ and $f''$, respectively. That programs are from the proof of Theorem \ref{th:deord-classic}.

Let us discuss complexity properties of $f^{(1)}$. Let set $\Sigma\subset\{0,1\}^n$ be a set of inputs allowed for $f'$. 
 Let $\pi$ be any order of variables, partition $\theta\in\Theta(\pi), \theta=(X_A,X_B)$ be such that $N^{\theta}(f')=N^{\pi}(f')$. 
If inputs $\nu,\nu'\in\{0,1\}^{|X_A|}$ such that corresponding subfunctions $f'|_\rho\neq f'|_{\rho'}$ then $f^{(1)}|_\rho\neq f^{(1)}|_{\rho'}$, because there is $\gamma\in\{0,1\}^{|X_B|}$ such that $(\nu,\gamma)\in\Sigma$, $f'|_\rho(\gamma)\neq f'|_{\rho'}(\gamma)$, hence $f^{(1)}|_\rho(\gamma)\neq f^{(1)}|_{\rho'}(\gamma)$. Therefore $N^{\theta}(f^{(1)})\geq N^{\pi}(f')$. It is right for any order and we have $N(f^{(1)})\geq N^{\pi}(f^{(1)})\geq N^{\pi}(f')$ and $N(f^{(1)})\geq N^{id}(f)$.

By the same way we can see that  $N(f^{(i)})\geq N^{id}(f)$, for $i\in\{2,\dots,7\}$.

Due to the definition of $f^{(i)}$, program $P'$ computes this function. And due to definitions of $f^{(i)}$, for $i\in\{2,\dots,7\}$ corresponding programs computing that functions. 

\section{Proof of Lemma \ref{lm:deq-dlower}}\label{apx:deq-dlower}
Let us consider any order $\pi$. Let us consider a set of inputs $\Xi\subset \{0,1\}^n$ such that $Adr(\xi,\pi(i))=i$ for $\xi\in\Xi$, $1\leq i \leq b$, $b= \lfloor n/(\lceil \log n \rceil +1) \rfloor$. Let us consider partition $\theta=(\{x_{\pi(1))},\dots,x_{\pi(u)}\},\{x_{\pi(u+1))},\dots,x_{\pi(n)}\}$, such that exactly $b/2$ {\em value} variables belong to $X_A$ and therefore exactly $b/2$ {\em value} variables belong to $X_B$.

Let us consider partition of input $\nu =(\sigma,\gamma)$ with respect to partition $\theta$. Let $\Sigma=\{\sigma\in\{0,1\}^{|X_A|}, (\sigma,\gamma)\in \Xi$ for some $\gamma\in\{0,1\}^{|X_B|}\}$, $\Gamma=\{\gamma\in\{0,1\}^{|X_B|}, (\sigma,\gamma)\in \Xi$ for some $\sigma\in\Sigma\}$. It is easy to see that $\Xi = \Sigma \times \Gamma$. Note that by definition of $\Xi$ for any $\sigma,\sigma'\in \Sigma$ if $\sigma\neq\sigma'$ then at least one of {\em value} bits is different in  these two strings.

Let us prove that for any $\sigma,\sigma'\in\Sigma$ such that $\sigma\neq\sigma$ and corresponding mappings $\rho$ and $\rho'$ we have $f|_\rho\neq f'|_\rho$. Let us consider $\gamma\in\Gamma$ such that {\em value} bits of $\sigma $ are equals to {\em value} bits of $\gamma$. Then $REQ(\sigma,\gamma)=1$ and $REQ(\sigma',\gamma)=0$ therefore $f|_\rho\neq f'|_\rho$. It means that $N^{\theta}(REQ)\geq |\Sigma|$. $|\Sigma|\geq 2^{b/2}\geq 2^{\lfloor n/(2\lceil \log n \rceil +2) \rfloor}$. $N^{\pi}(REQ)\geq N^{\theta}(REQ)$, $N(REQ)\geq N^{\pi}(REQ)$, because we have bound for any order $\pi$.
\section{Proof of Theorem \ref{th:quntum-hi}}\label{apx:prob-quant-hi}

It is easy to see that class for quantum OBDD of smaller width is subset of second one. Let us prove inequality of these pairs of classes.

 Due to Claim 2 of Lemma \ref{lm:deqb-modp-mwsb}, Boolean function $MOD_{2^{d/\delta}}\in${\bf BPOBDD}$_{d}$, at the same time for any bounded error quantum OBDD $P$, which computes the function, we have $w(P)\geq  O(d/\delta) >d/\delta^2$. Therefore $MOD_{2^{d/\delta}}\not\in${\bf BPOBDD}$_{d/\delta^2}$.

Due to Claim 1 of Lemma \ref{lm:deqb-modp-mwsb}, Boolean function $REQ_n^{d}\in${\bf BPOBDD}$_{d^2}$, at the same time for any bounded error quantum OBDD $P$, which computes the function, we have $w(P)\geq \lfloor d/\log_2 d \rfloor>d/\log^2_2 d$. Therefore $REQ_n^{d}\not\in${\bf BPOBDD}$_{d/\log^2_2 d}$.

Due to Claim 3 of Lemma \ref{lm:deqb-modp-mwsb}, Boolean function $MWS_n^{\log d}\in${\bf BQOBDD}$_d$, at the same time for any bounded error quantum OBDD $P$ we have $w(P)=2^{\Omega(\log d)}>2^{(\log d)/\delta}=d^{1/\delta}$. Therefore $MWS_n^{\log d}\not\in${\bf BQOBDD}$_{d^{1/\delta}}$.

\section{Proof of Lemma \ref{lm:dpj}}\label{apx:dpj}

Let us consider communication complexity of Boolean function. $t$-round communication protocol is communication game of two players $A$ and $B$ for computing Boolean function $f(X)$, first one knows only variables from subset $X_A\subset X$ and second one knows only variables from $X_B=X\backslash X_A$. After $t$ rounds of communication game the last of them knows result of function. We also consider probabilistic case of this protocol. And it will compute function with error $\varepsilon$ if probability of error is at most $\varepsilon$. For more detailed description see \cite{nw91}.

Let $t$-round communication protocol, which computes $f$ and $A$ starts communication game sends $r$ bits, then $C^{A,t}(f)=r$ is communication complexity of function $f$. $C^{B,t}(f)$ is the similar complexity, but for protocol, such that $B$ starts computation. $C^{A,t}_{\varepsilon}(f)$ and $C^{B,t}_{\varepsilon}(f)$ are similar complexities but for probabilistic protocol with error at most $\varepsilon$.

We have following results for $PJ_{k,n}$ function from \cite{nw91}:
$C^{B,t}(PJ_{t,n})=\Omega(\lfloor n/(2\lceil \log_2 n\rceil)\rfloor - t\log_2(\lfloor n/(2\lceil \log_2 n\rceil)\rfloor))$, $C^{B,t}_{1/3}(PJ_{t,n})=\Omega(\lfloor n/(2\lceil \log_2 n\rceil)\rfloor/k^2 - t\log_2(\lfloor n/(2\lceil \log_2 n\rceil)\rfloor))$ in case when we give $f_A$ to player A and $f_B$ to B.
\subsection{Proof of Claim 1}
Firstly, let us construct commutative $2k$-OBDD $P$ for $PJ_{2k-1,n}$.

Program $P$ reads variables in natural order and $w(P)=n^2$.
On the layer $i$ program $P$ computes $f^{(i)}(v_0)$. for $1\leq i \leq 2k-1$. Let us consider a level of program's layer. Each node of level corresponds to pair $(v,v')$, where $v$ is value of $f$ on previous iteration and $v'$ is value that we compute on current level, $1\leq v, v'\leq n$. Let $v=f^{(i-1)}(v_0)$, then computation on $i$-th layer starts on node $(v,0)$. After that program $P$ tests variable $x_j$ on some level and if $(v-1)\log n <j\leq v\log n$ then from the node $(v,v')$ $1$-edge leads to $(v, v'+2^{j-(v-1)\log_2 b}$ $mod$ $b)$ node of the next level and $0$-edge leads to $(v,v')$. Both edges from any other node $(v,v')$ leads to node $(v,v')$ of the next level. On the last level instead of leading to $(v,v')$ edge leads to $(v',0)$  of the first level of the next layer, if $i$ is even, and $(v'+b,0)$ otherwise.

Let us describe the last layer. It starts on $(v,0)$, for $v=f^{(2k-1)}(v_0)$. Then program $P$ tests variable $x_j$ on some level and if $(v-1)\log n <j\leq v\log n$ then from node $(v,v')$ edges lead to $(v, v' \oplus x_j)$ node of the next level. Both edges from any other node $(v,v')$ lead to node $(v,v')$ of the next level. On the last level for each node edge instead of leading to $(v,v')$ leads to $v'$-sink node.

It is easy to see that $P$ computes $PJ_{2k-1,n}$. The claim that $P$ is commutative follows from the commutative property of addition.

Secondly, we apply Theorem \ref{th:deord-classic} to the $P$ and get $2k$-OBDD $P'$ of width at most $n^3$, which computes partial version of $RPJ_{2k-1,n}$.

It is easy to see that $P'$ also computes total version of $RPJ_{2k-1,n}$.

\subsection{Proof of Claim 2}

Firstly, let us prove the claim for deterministic case. Assume that $RPJ_{2k-1,n}$ is computed by $k$-OBDD $P$ of width $w=2^{o(n/(k\log n)-\log (n/\log n))}$. $k$-OBDD $P$ can be simulated by $2k-1$-round communication protocol $R$, which sends at most $\lceil\log_2w\rceil(2k-1)$ bits. For prove this fact look, for example, at \cite{K16}. Let us consider only inputs from the set $\Sigma\subset\{0,1\}^n$ such that for $\sigma\in\Sigma$ we have $Adr(\sigma,i)=i+b$, for $1\leq i\leq b$ and $Adr(\sigma,i)=i-b$, for $b+1\leq i\leq 2b$. For these inputs our protocol will just compute $PJ_{2k-1,b}$, but $B$ starts computation in communication game. Therefore we can get protocol $R'$ such that $B$ starts computation, from protocol $R$, which computes  $PJ_{2k-1,b}$ and sends at most  $\lceil\log_2w\rceil(2k-1)$ bits. It means $C^{B,2k-1}(PJ_{2k-1,b})<o(n/\log n)=o(n/\log_2 n-(2k-1)\log_2 (n/\log_2 n))$, for $n=b\lceil \log{b}+1 \rceil$. This contradicts the claim $C^{B,t}(PJ_{t,b})=\Omega(b - t\log {b})$ from   \cite{nw91}.  

Secondly, let us to prove the claim for probabilistic case by the same way. Assume that $RPJ_{2k-1,n}$ is computed by bounded error $k$-POBDD $P$ of width $w=2^{o( n/(k^3\log n)  - \log(n/\log n))}$ with error probability at most $1/3$. Then by the same way we can show that $C^{B,2k-1}_{1/3}(PJ_{2k-1,b})<o(n/\log n)=o( n/(k^2\log n)  - (2k-1)\log(n/\log n))$.  This contradicts the claim $C^{B,t}_{1/3}(PJ_{t,b})=\Omega(b/k^2 - t\log{b})$ from   \cite{nw91}. 

\section{Proof of Theorem \ref{thm:hierarchy}}\label{apx:hierarchy}

\subsection{Proof of Claim 1}
By the definition we have P-$k$OBDD $\subseteq$ P-$2k$OBDD. Let us prove P-$k$OBDD $\neq$ P-$2k$OBDD.  

Let us consider $RPJ_{2k-1,n}$, then each $k$-OBDD computing the function has size  
$2^{\Omega(n/(k\log n)-\log (n/\log n))}\geq  
2^{\Omega(n/(n\log^{-3} n \log n)- \log (n/\log n))}=2^{\Omega(\log^2n)}
=n^{\Omega(\log n)}$, due to Lemma \ref{lm:dpj}.

 Therefore it has more than polynomial size. Hence $RPJ_{2k-1,n}\not\in$ P-$k$OBDD and $RPJ_{2k-1,n}\in$ P-$2k$OBDD, 
 due to Lemma \ref{lm:dpj}.
 
 By the definition we have BP-$k$OBDD $\subseteq$ BP-$2k$OBDD. Let us prove BP-$k$OBDD $\neq$ BP-$2k$OBDD.  

Let us consider $RPJ_{2k-1,n}$, then each $k$-POBDD computing the function has size  
$2^{\Omega( n/(k^3\log n)  - \log(n/\log n))}\geq  
2^{\Omega(n/(n\log^{-3} n \log n)- \log (n/\log n))}=2^{\Omega(\log^2n)}
=n^{\Omega(\log n)}$, due to Lemma \ref{lm:dpj}.

 Therefore it has more than polynomial size. Hence $RPJ_{2k-1,n}\not\in$ BP-$k$OBDD and $RPJ_{2k-1,n}\in$ BP-$2k$OBDD, 
 due to Lemma \ref{lm:dpj}.

\subsection{Proof of Claim 2}

By the definition we have SUPERPOLY-$k$OBDD $\subseteq$ SUPERPOLY-$2k$OBDD. Let us prove SUPERPOLY-$k$OBDD $\neq$ SUPERPOLY-$2k$OBDD.  

Let us consider $RPJ_{2k-1,n}$, then each $k$-OBDD computing the function has size  
$2^{\Omega(n/(k\log n)-\log (n/\log n))}\geq  
2^{\Omega(n/(n^{1-\delta} \log n) -\log (n/\log n))}=2^{\Omega(n^{\delta}/\log n -\log (n/\log n))}
>2^{\Omega(n^{\delta/2})}$, due to Lemma \ref{lm:dpj}.

 Therefore it has more than superpolynomial size. Hence $RPJ_{2k-1,n}\not\in$ SUPERPOLY-$k$OBDD and $RPJ_{2k-1,n}\in$ SUPERPOLY-$2k$OBDD, 
 due to Lemma \ref{lm:dpj}.

 By the definition we have BSUPERPOLY-$k$OBDD $\subseteq$ BSUPERPOLY-$2k$OBDD. Let us prove BSUPERPOLY-$k$OBDD $\neq$ BSUPERPOLY-$2k$OBDD.  

Let us consider $RPJ_{2k-1,n}$, then each $k$-POBDD computing the function has size  
$2^{\Omega( n/(k^3\log n)  - \log(n/\log n))}\geq  
2^{\Omega(n/(n^{1-3\delta} \log n) -\log (n/\log n))}=2^{\Omega(n^{3\delta}/\log n -\log (n/\log n))}
>2^{\Omega(n^{\delta})}$, due to Lemma \ref{lm:dpj}.

 Therefore it has more than superpolynomial size. Hence $RPJ_{2k-1,n}\not\in$ BSUPERPOLY-$k$OBDD and $RPJ_{2k-1,n}\in$ BSUPERPOLY-$2k$OBDD, 
 due to Lemma \ref{lm:dpj}.

\subsection{Proof of Claim 3}
By the definition we have SUBEXP$_{\alpha}$-$k$OBDD $\subseteq$ SUBEXP$_{\alpha}$-$2k$OBDD. Let us prove SUBEXP$_{\alpha}$-$k$OBDD $\neq$ SUBEXP$_{\alpha}$-$2k$OBDD.  

Let us consider $RPJ_{2k-1,n}$, then each $k$-OBDD computing the function has size  
$2^{\Omega(n/(k\log n)-\log (n/\log n))}\geq  
2^{\Omega(n/(n^{1-\delta} \log n)-\log (n/\log n))}=2^{\Omega(n^{\delta}/\log n)}
>2^{\Omega(n^{\delta-\varepsilon})}$, $2^{n^{\alpha}}=2^{o(n^{\delta-\varepsilon})}$, due to Lemma \ref{lm:dpj}.

 Therefore it has more than subexponential size $2^{\alpha}$. Hence $RPJ_{2k-1,n}\not\in$  SUBEXP$_{\alpha}$-$k$OBDD and $RPJ_{2k-1,n}\in$  SUBEXP$_{\alpha}$-$2k$OBDD, 
 due to Lemma \ref{lm:dpj}.

By the definition we have BSUBEXP$_{\alpha}$-$k$OBDD $\subseteq$ BSUBEXP$_{\alpha}$-$2k$OBDD. Let us prove BSUBEXP$_{\alpha}$-$k$OBDD $\neq$ BSUBEXP$_{\alpha}$-$2k$OBDD.  

Let us consider $RPJ_{2k-1,n}$, then each $k$-POBDD computing the function has size  
$2^{\Omega( n/(k^3\log n)  - \log(n/\log n))}\geq  
2^{\Omega(n/(n^{1-\delta} \log n)-\log (n/\log n))}=2^{\Omega(n^{\delta}/\log n)}
>2^{\Omega(n^{\delta-\varepsilon})}$, $2^{n^{\alpha}}=2^{o(n^{\delta-\varepsilon})}$, due to Lemma \ref{lm:dpj}.

 Therefore it has more than subexponential size $2^{\alpha}$. Hence $RPJ_{2k-1,n}\not\in$  BSUBEXP$_{\alpha}$-$k$OBDD and $RPJ_{2k-1,n}\in$  BSUBEXP$_{\alpha}$-$2k$OBDD, 
 due to Lemma \ref{lm:dpj}.
    

\end{document}